\documentclass[twocolumn]{article}
\usepackage[a4paper,left=1.5cm,right=1.5cm,top=3cm,bottom=4cm]{geometry}

\usepackage{amsmath,amssymb,amsfonts}   
\usepackage[bottom]{footmisc}

\usepackage{algorithmic}
\usepackage{hyperref}
\usepackage{graphicx}
\usepackage{subcaption}
\usepackage[sort&compress,super,square,comma]{natbib}
\usepackage{tabularx}
\usepackage{booktabs}
\usepackage{multirow}
\usepackage{float}
\usepackage{textcomp}
\usepackage{xcolor}
\usepackage[version=4]{mhchem}
\usepackage{enumerate}
\usepackage{enumitem}
\usepackage{siunitx}
\usepackage{upgreek}
\usepackage{authblk}

\usepackage{amsmath,amssymb}

\makeatletter
\newcommand{\@defs@vec}[1]{\boldsymbol{#1}}
\newcommand{\@defs@tens}[1]{\mathbf{#1}}

\newcommand{\fnorm}[1]{| #1 |\@defs@replaced{abs}} 

\newcommand{\vect}[1]{\@defs@vec{#1}{}}       
\newcommand{\tens}[1]{\@defs@tens{#1}{}}      

\newcommand{\nvec}{\@defs@vec{n}}

\newcommand{\SEI}{\text{SEI}}
\newcommand{\inte}{\text{int}}

\newcommand{\Lint}{{\text{e}^\text{-}}}
\newcommand{\LintO}{{\text{e}^\text{-},0}}

\newcommand{\cref}{c_\text{max}}

\makeatother

\makeatletter
\renewcommand*{\@fnsymbol}[1]{\ensuremath{\ifcase#1  \or = \or * \or \dagger\or \ddagger\or 
   \mathsection\or \mathparagraph\or * \or \|\or **\or \dagger\dagger
   \or \ddagger\ddagger \else\@ctrerr\fi}}
\makeatother

\begin{document}

\title{A four parameter model for the solid-electrolyte interphase to predict battery aging during operation}

\date{\today}

\author[1,2]{Lars~von~Kolzenberg \thanks{These authors contributed equally to this work}}
\author[3,4]{Jochen~Stadler  $^=$}
\author[4]{Johannes~Fath}
\author[4]{Madeleine~Ecker}
\author[1,2,3]{Birger~Horstmann}
\author[1,2,3]{Arnulf~Latz \thanks{Corresponding Author: arnulf.latz@dlr.de}}
\affil[1]{German Aerospace Center, Pfaffenwaldring 38-40, 70569 Stuttgart, Germany}
\affil[2]{Helmholtz Institute Ulm, Helmholtzstra{\ss}e 11, 89081 Ulm, Germany}
\affil[3]{Ulm University, Albert-Einstein-Allee 47, 89081 Ulm, Germany}
\affil[4]{Mercedes-Benz AG, Mercedesstr. 120, 70372 Stuttgart, Germany}

\maketitle

\begin{abstract}
Accurately predicting aging of lithium-ion batteries would help to prolong their lifespan, but remains a challenge owing to the complexity and interrelation of different aging mechanisms.
As a result, aging prediction often relies on empirical or data-driven approaches, which obtain their performance from analyzing large datasets.
However, these datasets are expensive to generate and the models are agnostic of the underlying physics and thus difficult to extrapolate to new conditions.
In this article, a physical model is used to predict capacity fade caused by solid-electrolyte interphase (SEI) growth in 62 automotive cells, aged with 28 different protocols. 
Three protocols parametrize the time, current and temperature dependence of the model, the state of charge dependence results from the anode's open circuit voltage curve. The model validation with the remaining 25 protocols shows a high predictivity with a root-mean squared error of $1.28\%$.
A case study with the so-validated model shows that the operating window, \textit{i.e.} maximum and minimum state of charge, has the largest impact on SEI growth, while the influence of the applied current is almost negligible.
Thereby the presented model is a promising approach to better understand, quantify and predict aging of lithium-ion batteries.
\end{abstract}

\section{Introduction}

Lithium-ion batteries are the current benchmark technology for mobile energy storage because they combine high energy density and longevity. Nevertheless,
different aging phenomena continuously decrease the usable capacity and limit the battery's lifetime. This is a major challenge for battery electric vehicles (BEV), which require a battery lifetime of about 10 years. To ensure a certain remaining capacity in this timespan, a detailed understanding and quantification of the capacity fade in lithium-ion batteries is imperative.

Two main approaches exist to predict capacity fade in lithium-ion batteries. On the one hand, (semi-)empirical approaches based on simplified physical equations or data driven methods generate a precise aging quantification from large datasets \cite{Kabitz2013,Schmalstieg2014,Groot2015,Hahn2018,Severson2019,Greenbank2021}. However, obtaining the necessary datasets is costly and the resulting models can not easily be extrapolated or adapted to new cell chemistries.
On the other hand, physics-based approaches model the underlying aging phenomena and infer equations to describe aging based on operating conditions \cite{Broussely2001,Ploehn2004,Christensen2004,Pinson2012,Tang2012,Li2015,Single2016,Single2017,Single2018,Das2019,VonKolzenberg2020}.

Most physical battery aging models rely on resolving solid-electrolyte interphase (SEI) growth as the major cause of continuous capacity fade \cite{fong1990studies,novak1999situ,naji1996electroreduction,Wang2018}. The SEI is a thin layer on the anode, which emerges in the initial battery cycle from electrochemical reactions of electrolyte molecules, electrons and lithium ions \cite{Peled1979,Peled1995,Peled1997,Aurbach1999,Aurbach2000,Winter2009,lu2014chemistry,Huang2019}.
In subsequent cycles, the SEI shields electrolyte molecules from electrons, but continues to grow due to a leak current through the SEI. 
 
During storage, the square-root-of-time ($\sqrt{t}$) dependent capacity fade points to a self-limiting kinetic behind this continuous growth process \cite{Broussely2001, Ploehn2004, Pinson2012, Single2018, Horstmann2018}. Diffusion of localized electrons, \textit{e.g.} as neutral lithium interstitial atoms \cite{Shi2012,Soto2015}, emerged as most likely growth limiting mechanism as it best captures the experimentally observed voltage dependence \cite{Keil2016,Keil2017,Single2018}.

During operation, recent experiments of Attia \textit{et al.} \cite{Attia2019} reveal that SEI growth accelerates with increasing charging current and decreases with increasing discharging current.
We orient us on these findings in our recent work \cite{VonKolzenberg2020} and extend the model of Single \textit{et al.} \cite{Single2018} for the effect of battery operation. The resulting model shows good accordance with both, storage and operation capacity fade measurements.

In this paper, we use the long-term limit of our model from ref. \citenum{VonKolzenberg2020} to describe capacity fade of 62 automotive grade pouch cells aged with 28 different protocols \cite{Stadler.2022}. Using differential voltage analysis (DVA), we first separate capacity fade into the three aging modes loss of active material on positive and negative electrode and loss of active lithium (LL), which we account to SEI growth \cite{Bloom.2005, Dubarry.2012}. Then, three protocols and the anode open circuit voltage (OCV) curve parametrize the time, current, temperature and state of charge (SoC) dependence of the model, while the remaining 25 protocols validate our model predictions.
Finally, post-mortem experiments with experimental coin cells built from harvested electrode materials provide further complementary validation for the model's assumptions. 

We present the experiments and the implemented model detailedly in the following Sections \ref{s:Experimental} and \ref{s:Theory}. In Section \ref{s:Results}, we present the experimental and theoretical results. Finally, Section \ref{s:Conclusion} summarizes the main findings of the work and shows future applications and extensions.

\section{Experimental}
\label{s:Experimental}

In the following section, we outline our experimental methods. Subsection \ref{ss:Aging_Test} concisely presents the cell specifications and the test matrix. Figure \ref{fig:DVAFitting} shows our methodology to separate capacity fade into different aging modes and isolate the influence of LL.
During operation, we determine LL with DVA, presented in Subsection \ref{ss:LL}.
Post mortem, we analyse heterogeneities using DVA as described in Subsection \ref{ss:Post_mortem}.

\subsection{Aging test}
\label{ss:Aging_Test}
The aging test comprises 62 automotive grade lithium ion pouch cells with a nominal capacity of $\SI{43}{\ampere\hour}$ and nominal voltage of $\SI{3.63}{\volt}$. The cells contain a graphite anode and a blend cathode consisting of $\ce{Li(Ni_{0.6}Mn_{0.2}Co_{0.2})O_2}$ and $\ce{Li(Ni_1/3Mn_1/3Co_1/3)O_2}$. The aging procedure is detailedly described in ref. \citenum{Stadler.2022} and the aging conditions are listed in Table SI-1.

In short, the cycling protocol consists of a charging sequence with a constant power $\text{P}_\text{CH}$ and subsequent constant voltage until reaching $\text{SoC}_\text{max}$. Afterwards, a dynamic EV profile discharges the cell to $\text{SoC}_\text{min}$ with maximum currents of up to $\SI{200}{\ampere}$ , followed by repetition of charge sustaining hybrid-EV profiles  until the specified $\text{EV}_\text{ratio}$, which is the ratio of energy throughput stemming from EV and hybrid-EV driving, is met.Real driving profiles are the basis for EV and hybrid-EV profiles which roughly compare to the widely used WLTP profiles.
To reveal the influence of these factors on aging, we use a central composite design of experiments. This approach varies the influencing factors around a common center point at $T=\SI{30}{\degreeCelsius}$, $\text{SoC}_\text{max}=90\%$, $\text{SoC}_\text{min} = 28\%$, $\text{EV}_\text{ratio} = 60\%$ and $\text{P}_\text{CH} = \SI{136}{\watt}$.
Additionally we perform one calendar aging test at $T=\SI{30}{\degreeCelsius}$ and $\text{SoC}_\text{max}=48\%$. 
Twofold measurement of each aging protocol and eightfold measurement of the center point results in 28 different conditions of the 62 cells, detailedly described in ref. \citenum{Stadler.2022} and listed in Table SI-1. 

\begin{figure*}[tb!]
 \centering
 \includegraphics[width=\textwidth]{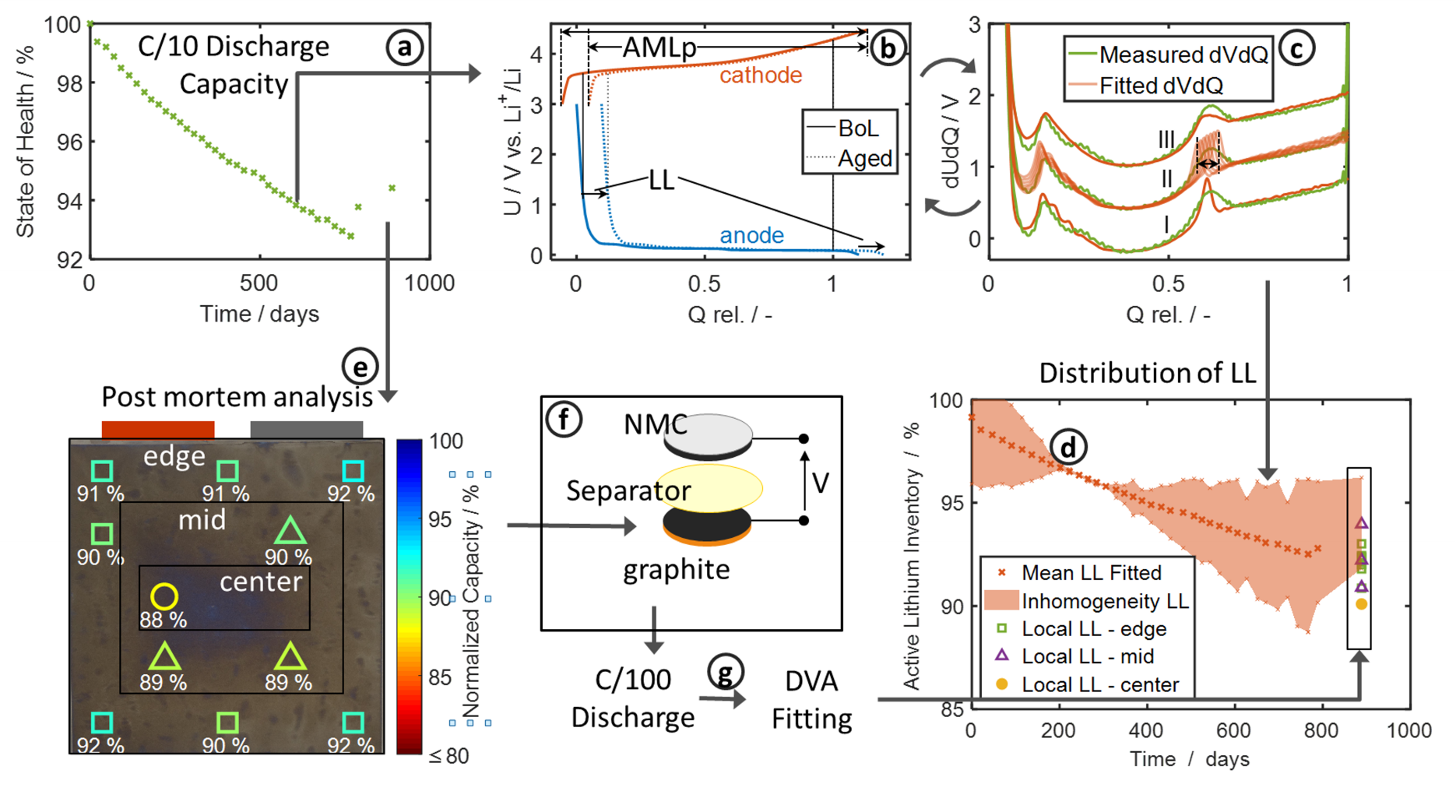}
 \caption{Evaluation of LL and homogeneity of lithium distribution. a) State of health determined by C/10 capacity check. b) The algorithm simulates aged dU/dQ measurements by shifting the electrode balancing as a result of LL and stinting the electrode potentials caused by loss of active material. The figure shows $10\%$ LL and $10\%$ active material loss on cathode side (AMLp). c) A superposition of simulated dU/dQ curves with a linear distribution of LL can reproduce the peak broadening in aged dU/dQ curves caused by inhomogeneous LL. d) Capacity fade due to lithium loss (crosses) and fitted distribution indicating inhomogeneities (orange area). d-g) During post mortem analysis experimental cells (f) are built with coins from harvested electrodes to determine local capacity loss within a layer (e). C/100 discharge curves are further used for DVA analysis (g) to validate local LL after the end of test (d).
   }
 \label{fig:DVAFitting}
\end{figure*}

To track the capacity fade we conduct reference parameter tests (RPT)  at $\SI{25}{\degreeCelsius}$ after every two weeks of cycling. These tests consist of three 1C/1C cycles and one 1C/0.1C cycle between \SI{4.2}{\volt} and \SI{2.5}{\volt} with CCCV charging with a cut-off at 0.05C and CC discharging. Additionally we measure $\SI{30}{\second}$ discharge pulses with a current of $\SI{200}{\ampere}$ at $\SI{100}{\percent}$, $\SI{75}{\percent}$, $\SI{50}{\percent}$ and $\SI{25}{\percent}$ SoC. For this work we focus on the C/10 capacity fade as exemplarily shown in Figure \ref{fig:DVAFitting}a).  In the context of the aging test a C-rate of 1C corresponds to the current that charges the cell within one hour based on a nominal capacity of \SI{43}{\ampere\hour}. In our analysis in Subsection \ref{ss:ModelAnalysis} we base the C-rate on the measured mean BoL capacity of \SI{45.89}{\ampere\hour}.
The measured capacity fade consists of an irreversible part and a reversible part originating from lateral flow of lithium into the anode overhang  \cite{Lewerenz.2019}. To retrieve the bound charge in the anode overhang and thus isolate the irreversible part, we remove  one cell for each aging condition from the aging test after about two years and discharge to $0\%$ SoC. Subsequently we perform another RPT after two to three months prior to post mortem analysis.

\subsection{Isolating active lithium loss}
\label{ss:LL}
We quantify the contribution of active material and active lithium loss on the overall capacity fade with the DVA fitting algorithm described by Fath \textit{et al.} \cite{Fath.2019} based on the C/10 discharge.
The algorithm relies on changes in the cell's voltage curve $U_\text{0,cell}=U_\text{0,cath}-U_\text{0,an}$ relative to the begin of life (BoL), see Figure \ref{fig:DVAFitting}b).
Loss of active material on the anode and cathode side stints the electrode potential curves $U_\text{0,cath}$ and $U_\text{0,an}$, respectively. In contrast, loss of lithium inventory, \textit{e.g.} stemming from SEI growth or lithium plating, shifts the anode's \textit{vs.} the cathode's potential curve \cite{Bloom.2005,Dubarry.2012,Fath.2019}.

The algorithm varies the BoL potential curves accordingly to fit the aged $\text{d}U/\text{d}Q$ curves and consists of two parts exemplarily shown in Figure \ref{fig:DVAFitting}c). First, the algorithm varies the LL and the loss of active material on anode and cathode side until the peak positions match, see I in Figure \ref{fig:DVAFitting}c).
Second, the algorithm fits the peak broadening observed in II-III in Figure \ref{fig:DVAFitting}c) as inhomogeneous aging by dividing the cell into seven parallelly connected cell segments. Each segment shows a particular LL according to a linear distribution around the previously determined mean LL \cite{Lewerenz.2018b,Sieg.2020}. 
From this distribution, the overall DVA results as weighted sum of the cell segments inverse $(\text{d}U/\text{d}Q)^{-1}$ \cite{Sieg.2020,Fath.2019,Dubarry.2019}. 
Both steps are repeated until the root mean square error (RMSE) between fitted and measured $\text{d}U/\text{d}Q$ curve is minimal. Minimum and maximum of the fitted LL then indicate the homogeneity of lithium distribution inside the cell.

 Fitting all available C/10 curves, we obtain a distribution of lithium loss (LL) across the cell throughout the entire course of aging, shown in Figure \ref{fig:DVAFitting} d). Here, the orange crosses indicate the capacity fade resulting from LL and the orange area shows the inhomogeneity of lithium loss based on the previously fitted distribution.
As we observe significant inhomogeneity after around two years, we also conducted post mortem experiments to validate our assumptions.

\subsection{Post mortem analysis}
\label{ss:Post_mortem}

As first step of our post mortem analysis, we uniformly discharge the cells with C/3 CCCV and C/50 cutoff to \SI{3}{\volt}. Then, we open the cells in a glovebox with argon atmosphere and $\ce{H_2O}$ and $\ce{O_2}$ levels below \SI{0.1}{ppm} following the procedure described by Sieg \textit{et al.} \cite{Sieg.2020}.
To remove residues of conducting salt, we wash anode and cathode layers from the middle of the stack in a bath of dimethyl carbonate (DMC) for at least three minutes. Afterwards, we remove active material from one side of the electrode and cut coins with \SI{18}{\milli\metre} diameter (see Fig \ref{fig:DVAFitting} d).
From coins that are opposing in the original pouch cell configuration, we assemble experimental cells using the PAT-Cell Setup from EL-Cell with the  FS-5P core cells according to Figure \ref{fig:DVAFitting} e). The cells contain \SI{95}{\micro\litre} of a DMC/EC electrolyte, a \SI{220}{\micro\metre} thick double layered PE-fibre / PP-membrane separator, a metallic lithium reference electrode, and aluminum and copper plungers as current collectors. Lastly, each cell undergoes a formation procedure  with CC cycles between \SI{4.2}{\volt} and \SI{2.5}{\volt}  consisting of 2x $\pm$ \SI{0.7}{\milli\ampere}, 2x $\pm$ \SI{1.4}{\milli\ampere}, 2x $\pm$ \SI{2.3}{\milli\ampere}, 1x $\pm$ \SI{0.7}{\milli\ampere}.

We electrochemically test the so-formed cells at \SI{25}{\degreeCelsius} with a BaSyTec CTS system. Discharge capacity and potential curves result from charging three times to 4.2V CCCV with \SI{5}{\milli\ampere} and \SI{0.07}{\milli\ampere} cutoff and subsequently CC discharging with \SI{0.07}{\milli\ampere} to \SI{2.5}{\volt}, corresponding to a C-rate of $\text{C}/100$ at BoL.

\section{Theory}
\label{s:Theory}

In this subsection, we present our mathematical model to describe capacity fade during battery operation caused by SEI growth. We start with a half-cell cycling model followed by an SEI growth model. Concludingly, we summarize the implemented system of equations.

\subsection{Half-Cell Cycling}

The applied intercalation current $J_\inte$ changes the half-cell charge 
$Q_\inte$ over time according to differential Equation \ref{eq:DE_Charging} 
\begin{equation}
\label{eq:DE_Charging}
\frac{\text{d}Q_\inte}{\text{d}t}=-J_\inte.
\end{equation}
Here, we use the IUPAC convention and define the intercalation current as negative and the deintercalation current as positive.  
We obtain the overpotential $\eta_\inte$ from the intercalation current $J_\inte$ based on a symmetric Butler-Volmer approach
\begin{equation}
\label{eq:Intercalation_Overpotential}
\eta_\inte=\frac{2RT}{F}\sinh^{-1}\left(\frac{J_\inte}{2J_\text{int,0}}\right)
\end{equation}
with Faraday's constant $F$, the universal gas constant $R$ and the temperature $T$ in Kelvin. The intercalation exchange current $J_\text{int,0}=J_\text{int,0,0}\sqrt{\tilde{c}}$ depends on the reaction rate $J_\text{int,0,0}$ and the SoC $\tilde{c}=Q_\inte/Q_\text{max}$ with the maximum capacity $Q_\text{max}$ \cite{Latz2013}. 
In the following we state equations to describe capacity fade resulting from SEI growth.

\subsection{SEI Growth}

We rely on our model developed in ref. \citenum{VonKolzenberg2020} to describe SEI growth during battery operation. From the multitude of possible reactions, we assume that the SEI predominantly forms from the reaction of lithium ions $\ce{Li}^+$, electrons $\ce{e^-}$ and electrolyte molecules, \textit{e.g.} ethylene carbonate $\ce{EC}$, according to reaction \ref{eq:SEI_formation_reaction}
\begin{equation}
\label{eq:SEI_formation_reaction}
2\ce{Li^+}+2\ce{e^-} +2\ce{EC} \rightarrow \ce{Li_2EDC} + \ce{R}.
\end{equation}
Here, $\ce{Li_2EDC}$ is the commonly observed SEI component lithium ethylene dicarbonate and $\ce{R}$ is a gaseous residue.
Assuming that lost lithium inventory $Q_\text{LL}$ is completely bound in the SEI with thickness $L_\SEI$, we linearly link both quantities with
\begin{equation}
    \label{eq:Irreversible_Capacity}
    Q_\text{LL} = \frac{\nu_\SEI A F}{\bar{V}_\SEI} L_\SEI.
\end{equation}
Here, $\nu_\SEI$ is the stoichiometric coefficient of lithium ions in the SEI formation reaction \ref{eq:SEI_formation_reaction}, $\bar{V}_\SEI$ is the mean molar volume of the resulting SEI components and $A$ denotes the active electrode surface.
Over time, the SEI grows in thickness according to differential Equation \ref{eq:SEI_Growth}
\begin{equation}
\label{eq:SEI_Growth}
\frac{\text{d}L_\SEI}{\text{d}t}=-\frac{\bar{V}_\SEI}{\nu_\SEI F}j_\SEI.
\end{equation}
The SEI formation current density $j_\SEI$ results from the diffusion of localized electrons, \textit{e.g.} as lithium interstitial atoms $\ce{Li^0}=\ce{Li^+}+\ce{e^-}$, from the electrode to the electrolyte \cite{Shi2012,Single2018,VonKolzenberg2020}. We describe this process with the long-term limit of our model developed in ref. \citenum{VonKolzenberg2020},
\begin{equation}
\label{eq:SEI_current_final}
j_\SEI=-\frac{c_\LintO D_\Lint F}{L_\SEI}e^{-\frac{F\eta_\SEI}{RT}}.
\end{equation}
Here, $c_\LintO$ and $D_\Lint$ are the equilibrium concentration and the diffusivity of lithium atoms inside the SEI. The diffusivity $D_\Lint$ depends on the temperature according to an Arrhenius kinetic
\begin{equation}
    \label{eq:Arrhenius}
    D_\Lint = D_{\LintO}e^{-\frac{E_\text{A}}{RT}}
\end{equation}
with the reference diffusivity $D_\LintO$ and the activation energy $E_\text{A}$.
The overpotential $\eta_\SEI$ for the formation of these lithium atoms depends on the OCV of the anode $U_0$ (see Figure SI-1), the intercalation overpotential $\eta_\inte$, and the reference chemical potential of lithium atoms $\mu_\LintO$, which we gauge to 0, according to
\begin{equation}
\label{eq:SEI_Overpotential}
\eta_\SEI = U_0 + \eta_\inte.
\end{equation}

\subsection{Model Summary}

In the following, we summarize our set of equations to determine active lithium loss due to SEI growth. We start with the differential equation to determine the SoC from the applied current profile
\begin{equation}
\label{eq:DE_Cycling}
    \frac{\text{d}\tilde{c}}{\text{d}t}=-\frac{J_\inte(t)}{Q_\text{max}}.
\end{equation}
From the SoC $\tilde{c}$ and the current profile $J_\inte(t)$, we then calculate the SEI formation overpotential
\begin{equation}
\label{eq:SEI_Overpotential:summary}
\eta_\SEI = U_0(\tilde{c}) + \frac{2RT}{F}\sinh^{-1}\left(\frac{J_\inte}{2J_\text{int,0,0}\sqrt{\tilde{c}}}\right).
\end{equation}
Followingly, we determine the SEI thickness evolution with the semi-analytical solution of Equation \ref{eq:SEI_Growth} and Equation \ref{eq:SEI_current_final}, Equation \ref{eq:SEI_Growth_Solution}
\begin{equation}
\label{eq:SEI_Growth_Solution}
    L_\SEI= \sqrt{2\frac{c_\LintO D_\LintO \bar{V}_\SEI}{\nu_\SEI}\int_0^te^{-\frac{F}{RT}\left(\eta_\SEI+\frac{E_\text{A}}{F}\right)}\text{d}t'+L_{\SEI,0}^2}.
\end{equation}
Lastly, the active lithium loss LL results from the SEI thickness \textit{via}
\begin{equation}
    \text{LL} = \frac{\nu_\SEI A F}{Q_\text{max} \bar{V}_\SEI} \left(L_\SEI-L_{\SEI,0} \right).
\end{equation}

Overall, the model relies on four fitting parameters to describe SEI growth. The diffusivity $D_\LintO$ and the initial SEI thickness $L_{\SEI,0}$ quantify the calendaric SEI growth rate. The exchange current density $J_\text{int,0,0}$ affects the resulting overpotential and thereby describes the current dependence of SEI growth. Lastly, the activation energy $E_\text{A}$ specifies the temperature dependence of SEI growth. 
Implementing the model equations \ref{eq:DE_Cycling}-\ref{eq:SEI_Growth_Solution} as numerical integration with \textit{e.g.} MATLABs \textit{trapz} yields results for yearlong cycling protocols within seconds.

\section{Results and Validation}
\label{s:Results}

In this section, we present the experimental and theoretical results obtained from the precedingly discussed experiments and model. We start by pointing out the dominant influence of lithium loss on capacity fade which we analyze with the DVA. Afterwards we compare the measured lithium loss of the test cells to our theoretical predictions and thereby parametrize and validate our model. Subsequently, a post mortem analysis shows the validity of our experimental and theoretical assumptions. Concludingly, we show the sensitivity of our SEI growth model to operating patterns and identify detrimental operating conditions.

\subsection{Identifying the cause of capacity fade}

Using the DVA fitting algorithm as presented in Subsection \ref{ss:LL} we quantify the contribution of the aging modes loss of active lithium (LL) and active material losses on anode side (AMLn) and on cathode side (AMLp) to overall capacity loss.

\begin{figure*}[htb]
 \centering
 \includegraphics[width=15 cm]{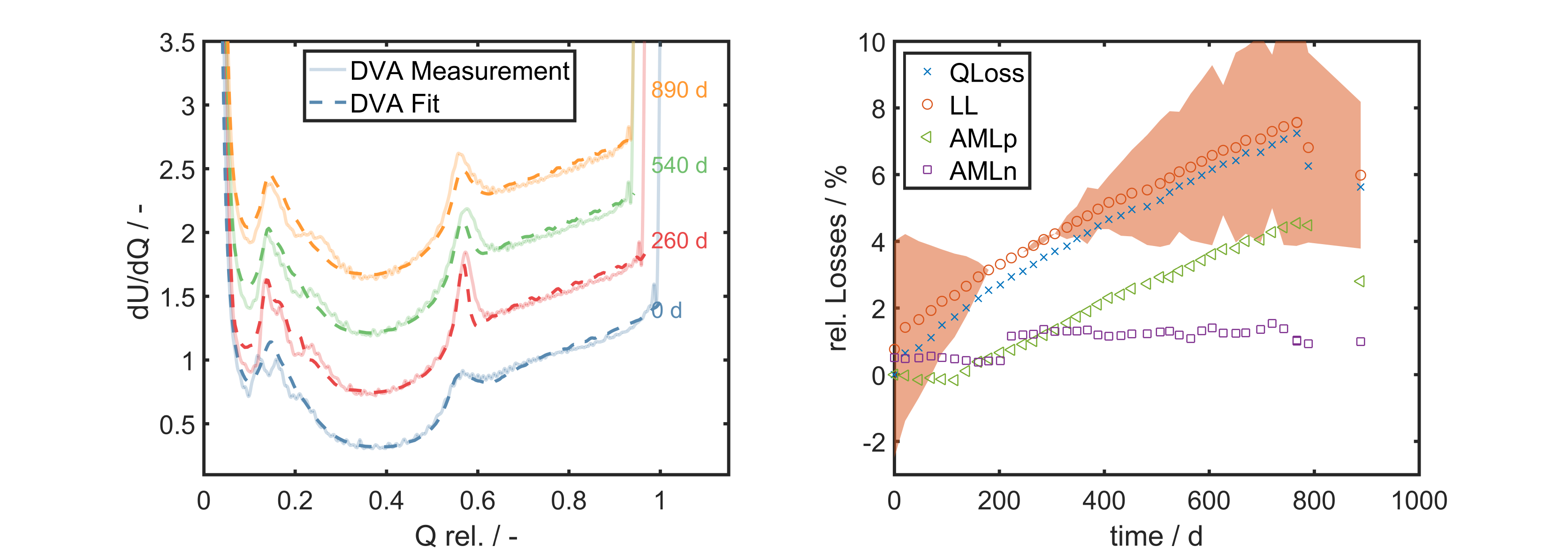}
 \caption{a) Measured and fitted DVA with increasing testing time from bottom to top. The peaks do not significantly shift which  b) Measured capacity fade QLoss in comparison to fitted loss of active lithium LL as well as active material losses on anode side (AMLn) and cathode side (AMLp). The orange area indicates inhomogeneity of lithium loss.}
 \label{fig:DVAresults}
\end{figure*}

Figure \ref{fig:DVAresults}a) exemplarily shows measured and fitted dU/dQ curves at different times for the reference cell 39. In the measured curves we observe only slight changes in the peak positions, pointing to negligible active material losses \cite{Dubarry.2012}. 
In the beginning, the peaks become sharper, which indicates a more homogeneous lithium distribution \cite{Sieg.2020}, possibly due to the anode overhang effect\cite{Fath.2020,Lewerenz.2019}. After about 300 days the peaks broaden again, pointing towards a more heterogeneous lithium distribution due to heterogeneous aging, as  we also observe in our post mortem analysis conducted after ending of test, see Section \ref{ss:Post_Mortem_Results}.

The results of the fitting algorithm in Figure \ref{fig:DVAFitting}b) support these observations and further show that AMLp exists but is not limiting.
The fitted LL shows good accordance with the measured capacity loss except for a slight offset at BoL, which we attribute to a slight mismatch in initial balancing due to inhomogeneities.

We emphasize that overall capacity loss is not equal to the sum of LL, AMLp and AMLn. Instead, capacity loss stems from limitations at the lower or upper cutoff voltage, which in turn result from changes in electrode balancing due to the different aging modes.
For example, the aging simulation with AMLp and LL in Figure \ref{fig:DVAFitting}b) shows, that LL shifts the steep anode potential and thus limits the lower cutoff potential, while AMLp is not significantly limiting the lower cutoff potential as it only slightly shifts the shallow cathode potential.

We observe similar trends with LL being the dominant aging factor for all cells. This allows us to focus our modeling approach on the main aging mechanism SEI growth as root-cause of LL and thus to keep the model simple.
However, harsher aging conditions or different cell chemistries can induce more severe active material losses, where our simple approach might fall short.

\subsection{Predicting lithium loss}

\subsubsection{Parametrization}

\begin{figure}[tb]
 \centering
 \includegraphics[width=8.4 cm]{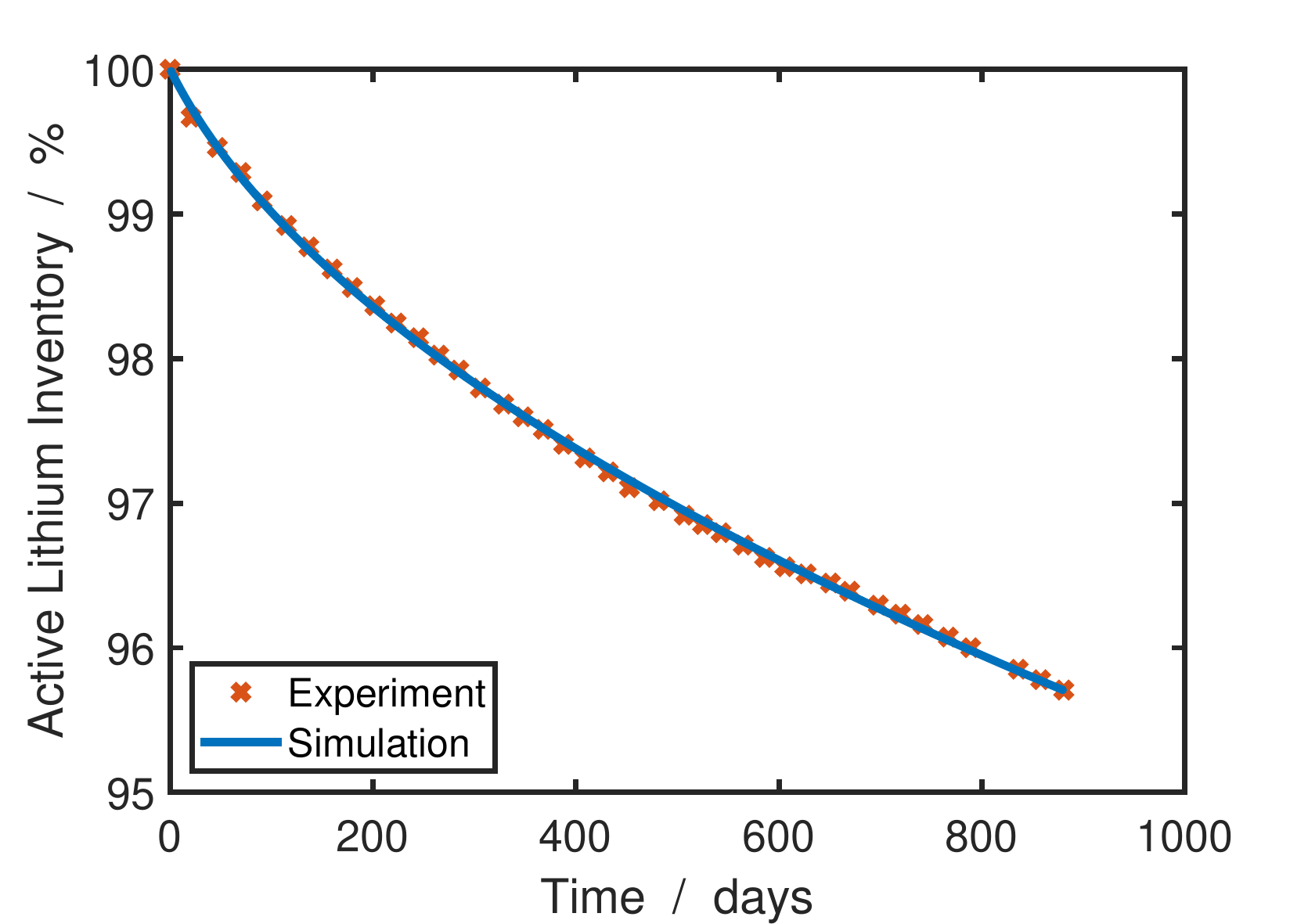}
 \caption{Comparison of experimental (orange crosses) and simulated (blue line) capacity fade during storage at $T=\SI{30}{\celsius}$ with $50\%$ SoC.}
 \label{fig:StorageParametrization}
\end{figure}

As a first step, we use storage data to parametrize the initial SEI thickness $L_{\SEI,0}$ and the diffusivity $D_\Lint(T=\SI{30}{\celsius})$, see Equation \ref{eq:Arrhenius}. In this case, we solve Equation \ref{eq:SEI_Growth_Solution} analytically, because the SEI overpotential $\eta_\SEI$ is constant in time. This leads to the well-known $\sqrt{t}$-dependence of SEI growth during battery storage \cite{Single2018}
\begin{equation}
    \label{eq:Storage_Solution}
    L_\SEI= \sqrt{2\frac{c_\LintO D_\Lint(T) \bar{V}_\SEI}{\nu_\SEI}e^{-\frac{F}{RT}U_0}t+L_{\SEI,0}^2}.
\end{equation} 
Figure \ref{fig:StorageParametrization} compares the remaining active lithium inventory predicted by our model with experimental data.
We see that the analytically derived $\sqrt{t}$ SEI growth law, Equation \ref{eq:Storage_Solution}, agrees excellently with the experimentally obtained capacity fade curves. In the following, we thus rely on the so-determined values for $L_{\SEI,0}$ and $D_\Lint(T=\SI{30}{\celsius})$ and proceed to determine the two remaining fitting parameters.

Next, we parametrize the influence of charging current on SEI growth with the exchange current $J_{\inte,0,0}$. For parametrization, we choose cell 39 from the reference cells 39-47, which were cycled at the same temperature as the storage cell and showed low variance among the eight fold measurement. 
\begin{figure*}[htb]
 \centering
 \includegraphics[width=\textwidth]{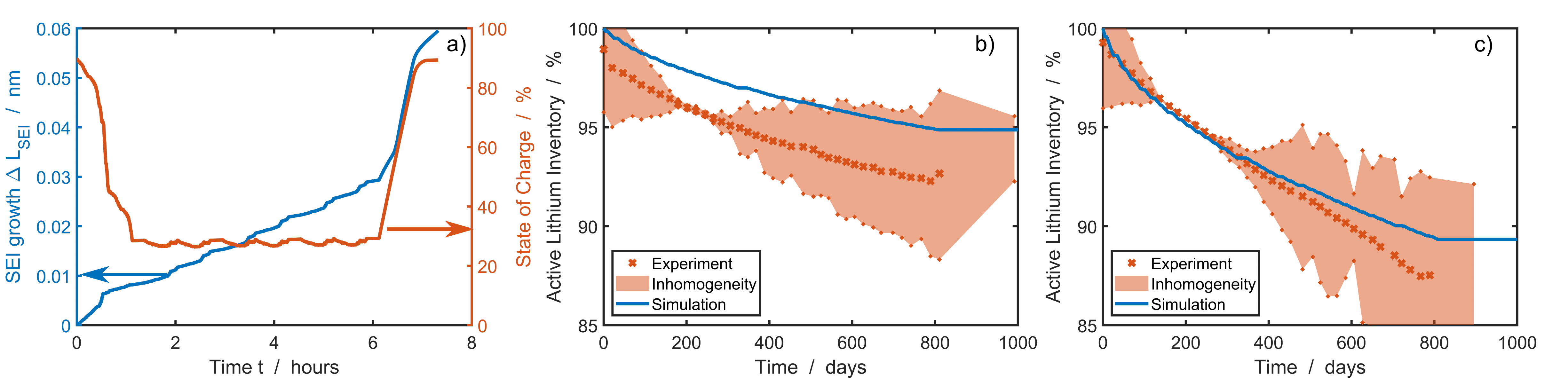}
 \caption{a) SEI growth simulation for the current profile of the reference cell 39 (see Table SI-1), which was continuously repeated in the aging protocol. The blue curve shows the SEI growth during the cycle based on an initial SEI thickness $L_{\SEI,0}=\SI{3.9}{\nano\meter}$, the orange curve shows the SoC.
 Comparison of experimental (orange crosses) and simulated (blue line) fade of active lithium inventory for b) the reference cell 39 cycled with the protocol listed in Table SI-1 at $T=\SI{30}{\celsius}$ and c) the high temperature cell 19 which were cycled with a distinct current profile, see Table SI-1 at $T=\SI{50}{\celsius}$.
 The orange areas indicate  heterogeneity within the cell as measured by DVA.}
 \label{fig:SEIGrowth39}
\end{figure*}
Before comparing our simulations with experiments, we illustrate the simulated SoC profile and the resulting SEI growth in Figure \ref{fig:SEIGrowth39} a). The orange line indicates the SoC of the battery during one cycle profile, which was then repeated several times. For the center cells, the SoC was mostly kept around 30\% with a charging/discharging sequence to 90\% in between. 
This protocol largely influences the SEI growth, which is shown in the blue curve. We observe that SEI growth proceeds slowly in the 30\% SoC phases. During charging to 90\%, the SEI growth is fastest, followed by the storage at 90\%. 
This accelerated growth results from the influence of intercalation overpotential $\eta_\inte$ and OCV $U_0$ on the SEI overpotential $\eta_\SEI$ according to Equation \ref{eq:SEI_Overpotential}, which is in line with previous capacity fade measurements \cite{Keil2016,Keil2017,Attia2019}.

Figure \ref{fig:SEIGrowth39}b) compares the average lithium loss in the center point cells with the simulated capacity fade.
The distribution of LL (orange area) exhibits a hourglass shape. In the first 150 days lithium distribution becomes more homogeneous, possibly due to the anode overhang effect. After about 300 days lithium distribution becomes increasingly inhomogeneous due to heterogeneous aging, as  we also observe in our post mortem analysis conducted after ending of test, see Section \ref{ss:Post_Mortem_Results}.

Overall, our mathematical SEI model agrees well with the experiments, especially in the homogeneous area between 200 and 400 days.
However, between 0 and 100 days, we observe a deviation of experiment and simulation, primarily because the experimental active lithium inventory starts below 100\%. We attribute this offset to experimental uncertainties, as the BoL electrode potentials that were used as basis for DVA fitting were measured well after test start. Hence the BoL cell can stem from a different production batch and thus show different initial lithium distribution resulting \textit{e.g.} from the reversible anode overhang effect \cite{Fath.2020,Lewerenz.2019}.
The kinks in the simulation profile result from the modular build-up of our simulation consisting of storage, cycle-profile and RPT phases. As each of these phases shows a different aging behaviour, we observe kinks between the profiles. After 800 days the active lithium inventory remains nearly constant in our simulation, because the battery is stored at a low SoC.

The intercalation reaction rate $J_{\inte,0,0}$ is not exclusive to our SEI model, but also a parameter in the commonly used Doyle-Fuller-Newman battery model \cite{Fuller1994}. This enables a comparison of our value $j_{\inte,0,0}=J_{\inte,0,0}/A=\SI{0.09}{\ampere\per\square\meter}$ with values found in literature. Sauer and coworkers \cite{Ecker2015,Schmalstieg2018} derived values ranging from $\SI{0.7}{\ampere\per\square\meter}$ to $\SI{7}{\ampere\per\square\meter}$ from electrochemical impedance spectroscopy (EIS). However, the determination of $j_{\inte,0,0}$ is ambiguous as the electrode active surface area $A$ is not easily attainable. As a consequence, Ng \textit{et al.} \cite{Ng2020} determine the exchange current of a $\SI{50}{\ampere\hour}$ cell with galvanostatic intermittent titration technique (GITT) to be around $\SI{10}{\ampere}$, which is comparable to our value of $J_{\inte,0,0}=\SI{21}{\ampere}$ for a $\SI{46}{\ampere\hour}$ cell.

Building up on this parametrization, we next determine the dependence of SEI growth on temperature by fitting the diffusivity $D_\Lint(T=\SI{50}{\celsius})$ to the high temperature cells cycled at $T=\SI{50}{\celsius}$. The parameters $D_\LintO$ as well as $E_\text{A}$ then result from the Arrhenius Equation \ref{eq:Arrhenius} and the diffusivity $D_\Lint(T=\SI{30}{\celsius})$, which we previously determined from storage data, see Figure \ref{fig:StorageParametrization}.
In Figure \ref{fig:SEIGrowth39}c), we again observe the hourglass shape of heterogeneity with a minimum around 200-400 days. Compared to the previous cell, the heterogeneity is even more pronounced in the long term and spans up to 10\% at around 800 days.
Our model shows good accordance to the average experimental values, but deviates in the long term.
As we also observe the highest heterogeneity in this case, we account the long term deviation between simulation and experiment to additional thermal effects, which are not captured by our simple SEI growth model. These effects could comprise thermal SEI degradation, lithium plating or electrolyte dry out. 

\subsubsection{Validation}

In this section, we validate our aging model with the remaining 25 protocols, which were not used for parametrization.
\begin{figure}[tb!]
 \centering
 \includegraphics[width=8.4 cm]{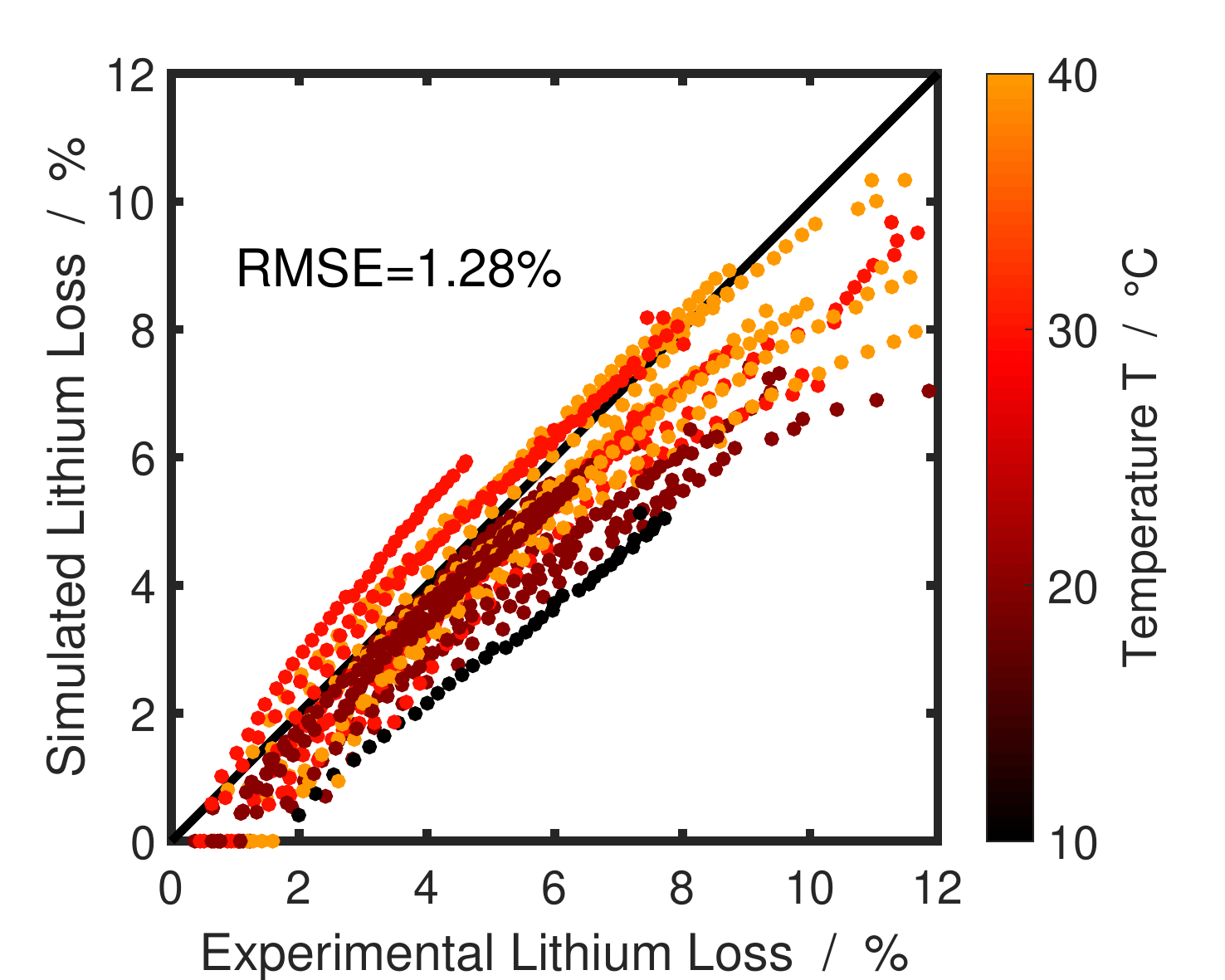}
 \caption{Scatterplot of the simulated active lithium loss $\text{LL}_\text{sim}$ \textit{vs.} the experimentally measured active lithium loss $\text{LL}_\text{exp}$}
 \label{fig:Scatterplot}
\end{figure}
Figure \ref{fig:Scatterplot} summarizes the validation results in a scatterplot, which compares the simulated and the experimentally measured lithium loss. The complete results of our aging simulations are listed in Figure SI-4.
The points are colored according to the operating temperature of the cells, which ranged from $\SI{10}{\celsius}$ to $\SI{40}{\celsius}$. In general, we observe a good accordance of simulation and experiments, which also reflects in the low root-mean-square error of $\text{RMSE}=1.28\%$. This underlines the outstanding global predicitivity of our model, which we obtained already after parametrizing the model with only three aging protocols.

However, we observe a systematic deviation of our simulated results from the experiments for temperatures below $T=\SI{20}{\celsius}$ as well as high aging states with more than 10\% lithium loss. 
We attribute this deviation to increased lithium plating at low temperatures as well as the large heterogeneity observed for these cells. Also the mentioned offset of measured LL at beginning of test amplifies the deviation. Followingly, we further validate these assumptions with post mortem analysis.

\subsection{Post Mortem Validation}
\label{ss:Post_Mortem_Results}

\begin{figure*}[tb!]
 \centering
 \includegraphics[width=\textwidth]{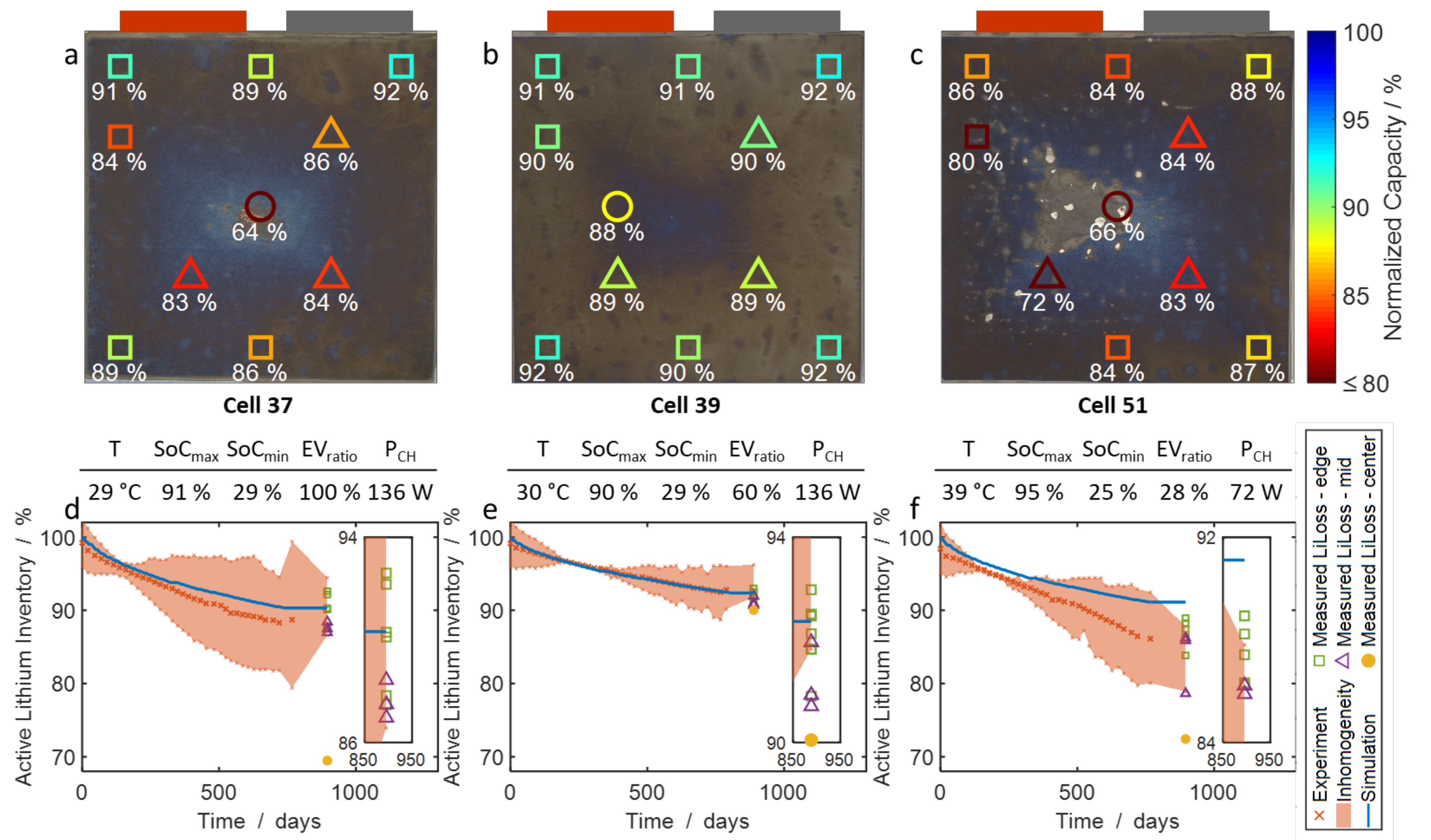}
 \caption{a-c) Photographs of anode electrode layer taken from the stack and respective local capacities of experimental cells in colored relation to BoL. Photographs reveal a moderate aging over the entire electrode for Cell 39 and severe aging and plating mainly focused to the center of the electrode for Cells 37 and 51.
 d-f) Measured and simulated loss of lithium inventory for the three exemplary cells. Orange indicates the experimental results consisting of an average active lithium inventory (cross) and inhomogeneity (orange area). The blue line shows the simulated loss of active lithium caused by SEI-growth. Markers show the active lithium loss fitted to the local three electrode cells. }
 \label{fig:ExperimentalResults}
\end{figure*}

In Figure \ref{fig:ExperimentalResults}, we exemplarily illustrate the model predictions for three cells cycled with different protocols at different temperatures and compare them to the measured local LL from experimental cells prepared as part of the post mortem analysis.

Visually, we already suspect an accelerated degradation focused to the center of the electrode from the anode photographs of the three cells in Figure \ref{fig:ExperimentalResults} due to the intensified blue coloring. Additionally cell 37 and 51 also exhibit shiny metallic grey spots that we attribute to lithium plating. Experimental cell measurement of particular anode areas allow to further evaluate heterogeneous capacity fading. For this purpose, we separate the electrode area into edge ($\square$), mid($\triangle$) and center ($\bigcirc$) as indicated in Figure \ref{fig:DVAFitting}. Local LL of the respective areas are shown as points at the end of life in the bottom part plots of Figure \ref{fig:ExperimentalResults}.

We observe that especially the center shows the highest LL and thus drives inhomogeneity of lithium distribution for all three cells. In contrast, the edge and mid areas age more homogeneously and our model nicely accords to this experimental LL for the moderately aged cells shown in Figure \ref{fig:ExperimentalResults} d) and e).
However, we see a large deviation to the fitted mean LL for the more harshly aged cell 51, Figure \ref{fig:ExperimentalResults} f). We attribute this model deviation to increased aging due to lithium plating in this case, as is also evident from the photograph in Figure \ref{fig:ExperimentalResults} c). 
To further increase our model accuracy for these cases, inhomogeneous SEI growth and lithium plating should be considered in future works.

\subsection{Identifying detrimental conditions}
\label{ss:ModelAnalysis}

In this section, we conduct a sensitivity analysis to unravel particularly detrimental operating patterns. Subsequently, we study the impact of different, simplified real-life motivated user profiles on aging in a case study.

\subsubsection{Sensitivity analysis}

\begin{figure}[tb!]
 \centering
 \includegraphics[width=7 cm]{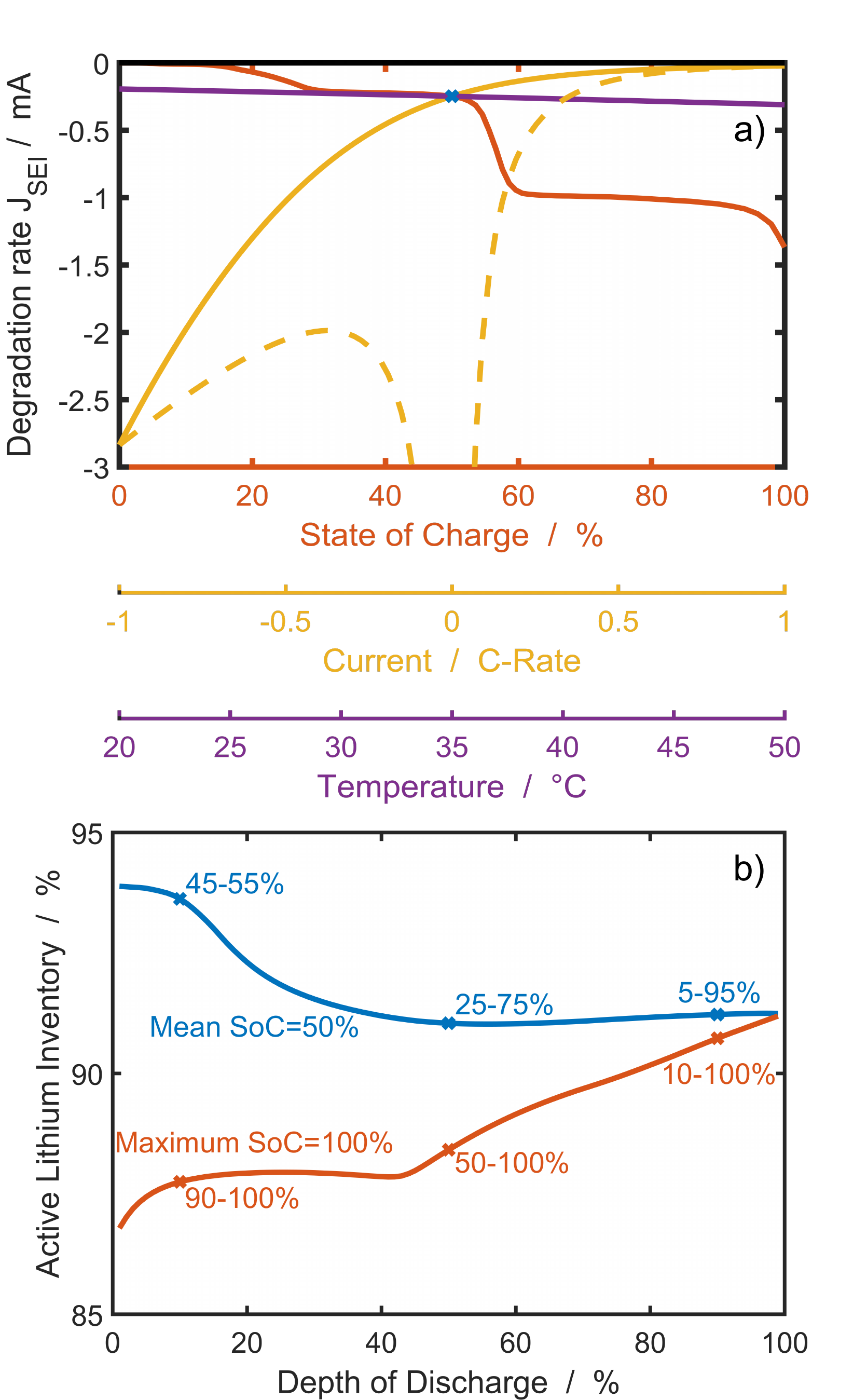}
 \caption{a) Sensitivity analysis of the influence of current, SoC and temperature on the SEI growth rate. The blue cross indicates the center point $J_\inte=0$, $T=\SI{35}{\celsius}$, $\text{SoC}=50\%$ around which we vary one factor at a time. Orange indicates the influence of the SoC, yellow the influence of the charging/discharging current and purple the influence of the temperature. The dashed yellow line shows the normalized degradation rate $\bar{J}_\SEI=J_\SEI\cdot t_\text{cycle}/\SI{1}{\hour}$.
 b) Sensitivity analysis of the depth of discharge influence (maximum SoC - minimum SoC) on the active lithium loss after cycling with 1C for two years. The blue curve shows the active lithium loss for DoD variation around a mean SoC=50\% whereas the red curve shows for the effect of DoD variation from a maximum SoC=100\%.}
 \label{fig:Sensitivity}
\end{figure}
Figure \ref{fig:Sensitivity}a) shows the influence of current $J_{\inte}$, SoC $\tilde{c}$ and temperature $T$ on aging compared to a central point, which is stored ($J_{\inte}=0$) with $\text{SoC}=50\%$ at $T=\SI{35}{\celsius}$. We observe that the influence of electrochemistry, \textit{i.e.} applied current and SoC exceeds the influence of temperature.
In particular, we observe the highest SEI current $J_\SEI$ for charging at a rate of 1C. This influence arises from the intercalation overpotential $\eta_\inte$, which increases the SEI overpotential $\eta_\SEI$ (Equation \ref{eq:SEI_Overpotential}) and thereby accelerates SEI growth. During discharging, in contrast, increasing the C-Rate has the adverse effect and decelerates the SEI growth until it is nearly suppressed at discharging with 1C. This is in good accordance with recent experiments of Attia \textit{et al.} \cite{Attia2019}, who showed this asymmetry in SEI growth between charging and discharging and thus motivated our model extension in reference \cite{VonKolzenberg2020}.

However, the shown influence of applied current on degradation is misleading, because the applied current also affects the cycle time. To emphasize this relationship, we also plot the normalized degradation rate $\bar{J}_\SEI=J_\SEI\cdot t_\text{cycle}/\SI{1}{\hour}/$ as dashed yellow line in Figure \ref{fig:Sensitivity}a). This rate compares the mean cumulated SEI growth within one cycle.
We observe three main effects: First, the influence of charging current on degradation is not as severe as we would expect solely from the yellow graph. Second, the degradation per cycle diverges for $J_\text{int}\rightarrow 0$, because the long cycle times increase the calendaric aging. Third, the opposing effects of current and cycle time on the degradation rate lead to an optimum at around C/3. To further analyse the complex interrelation of charging current and cycle time, we conduct a case study with different charging profiles in Subsection \ref{ss:CaseStudy}.

Another major contribution to SEI growth arises from the state of charge. We observe in Figure \ref{fig:Sensitivity}a) that the SEI current closely follows the anode OCV-curve (see Figure SI-1), in line with the experimental results of Keil \textit{et al.} \cite{Keil2016,Keil2017} and the model of Single \textit{et al.} \cite{Single2018}. This dependence causes cells stored at 60\% SoC to age nearly four times as fast as cells stored at 50\% SoC, while storage at 80\% leads to a similar capacity fade as storage at 60\%.


This trend also reflects in the depth of discharge $\text{DoD}=\text{SoC}_\text{max}-\text{SoC}_\text{min}$, which we analyze in the following.
Figure \ref{fig:Sensitivity}b) plots the remaining active lithium inventory after cycling for two years with $\text{J}_\inte=\pm 1\text{C}$, depending on the DoD. We simulate two different approaches; The blue curve shows a DoD variation around a mean SoC=50\%, while the orange curve varies the DoD with a fixed maximum SoC=100\%.

The blue curve shows that increasing the DoD around a mean SoC causes accelerated aging up to a plateau of 91\% active lithium inventory at 50\% DoD. This dependence results from the SoC-dependence of aging, see Figure \ref{fig:Sensitivity}a). In the regime of 0-50\% DoD, the battery experiences increasing share of SoCs larger than 60\%, which according to Figure \ref{fig:Sensitivity}a) increases aging. In the regime of 50-100\% this trend is balanced by the decreased aging for SoCs below 20\% leading to the observable plateau. 
In comparable experiments, Ecker \textit{et al.} \cite{Ecker.2014}  and Hoog \textit{et al.} \cite{Hoog.2017} observed an approximately linear dependency of capacity degradation with increasing DoD when cycling  NMC/graphite cells around a mean SoC of 50\%. Contradictory Sarasketa-Zabala \textit{et al.} \cite{SarasketaZabala.2015} showed a more complex dependency for LFP/graphite cells with the highest capacity fade between 10\% and 50\% DoD and lower capacity fade at very high and very small DoD, again while cycling around a mean SoC of 50\%.

The orange curve shows that for cycling around a fixed maximum SoC of 100\%, increasing the DoD results in slower aging. Again, this trend originates from the influence of SoC on aging, depicted in Figure \ref{fig:Sensitivity}a).
A systematic DoD variation from 100\% maximum SoC is not reported as often in literature, but results from Rechkemmer \textit{et al.} \cite{Rechkemmer.2020} for $\ce{LiMn_2O}$  hint to accelerated degradation at high DoD which is in contrast to our predictions based on anode SoC. Also Laresgoiti \textit{et al.} \cite{Laresgoiti.2015} derive an exponential acceleration of SEI-growth with increasing DoD from their SEI cracking approach as the particles undergo greater volume change. Benavente-Araoz \textit{et al.} \cite{BenaventeAraoz.2020} on the other hand observed a significantly higher degradation for cycling between 65-95\% SoC compared to 20-95 \% which is qualitatively in line with the SoC dependence as predicted in our model. 

Summing up the depth of discharge dependence of aging predicted by our model compared to literature \cite{Ecker.2014,Hoog.2017,SarasketaZabala.2015,Rechkemmer.2020,Laresgoiti.2015,BenaventeAraoz.2020}, we at most partially reproduce the experimentally observed trends. Accordingly, the influence of DoD on aging arises not only from a shift of the anode OCV, but also from additional effects. For example, the DoD affects particle volume changes and thus may cause SEI fracture or loss of electrical contact of individual particles.
To capture these effects and increase our model predictivity in this respect, extending the present simplistic SEI model for mechanics seems promising \cite{Castelli2021,VonKolzenberg2021}.

\subsubsection{Case Study}
\label{ss:CaseStudy}

We now apply our model to study battery aging in different, real-life motivated battery user profiles. We consider three electric vehicle enthusiasts: standard driver Stan (blue), range anxious Robert (red) and aging optimizer Alfred (yellow), who travel $\SI{50}{\kilo\meter}$ to work back and forth every working day. All three drive steadily at $\SI{50}{\kilo\meter\per\hour}$ and for simplicity discharge their EV with a range of $\SI{500}{\kilo\meter}$ constantly at C/10 for one hour at a constant temperature of $\SI{20}{\degreeCelsius}$. 
\begin{figure}[tb!]
         \centering
         \includegraphics[width=8.4cm]{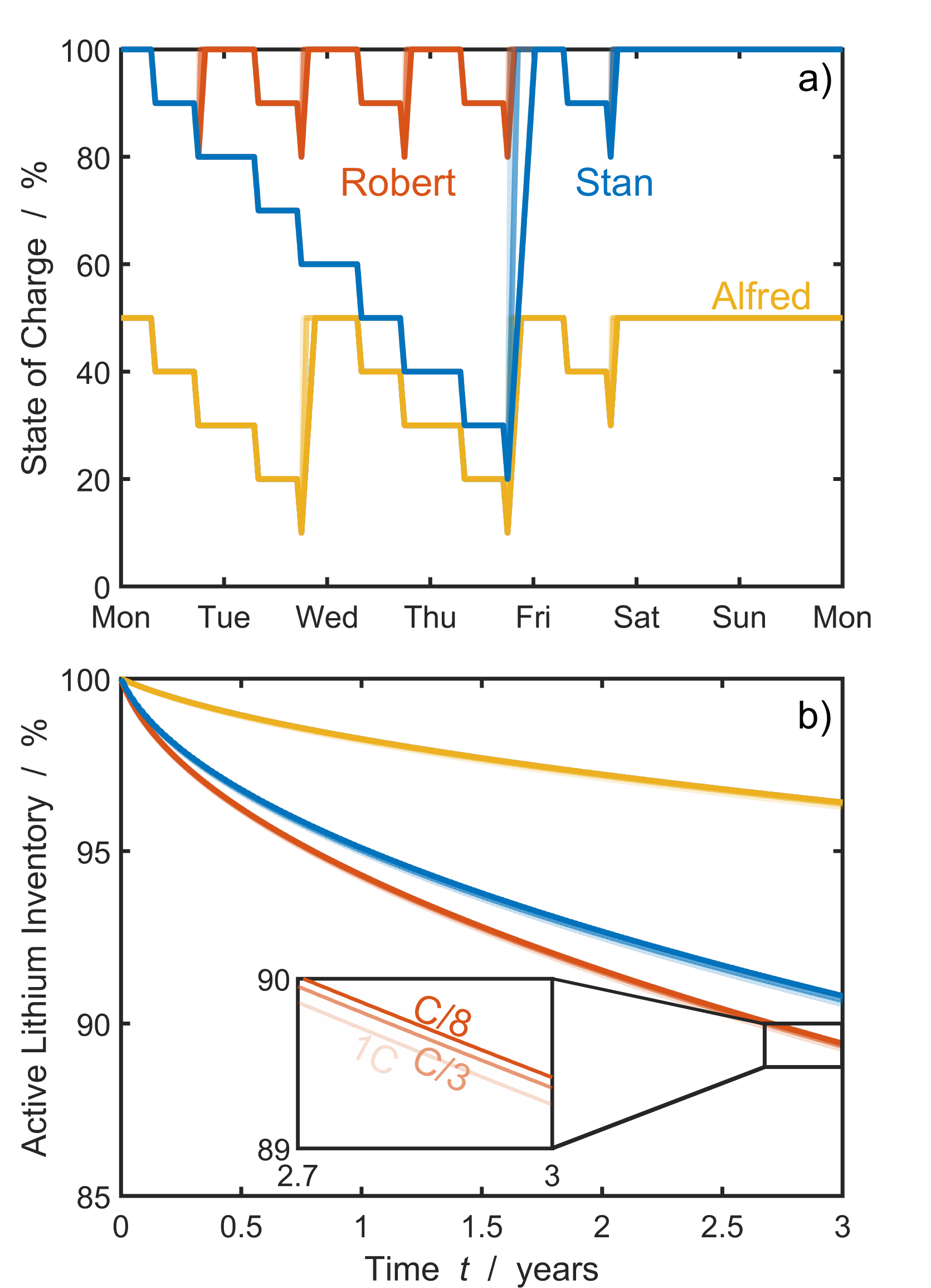}
 \caption{Aging results for different exemplary battery use cases at a temperature $T=\SI{20}{\celsius}$. a) State of charge over the course of a working week. During the working days, the battery is discharged twofold with C/10 for 10\% SoC in each case. Different colors indicate different charging behaviours. Blue: Standard behaviour with charging once the SoC goes beyond 40\% SoC. Red: Range anxiety with charging once the SoC is below 100\%. Yellow: Aging optimization with charging once the SoC is below 20\% SoC and recharging up to 50\% SoC. The standard charging protocol charges with C/8, C/3 is colored brighter and 1C the brightest. b) Battery aging after applying the user profile of a) continuously for three years.}
 \label{fig:CaseStudy}
\end{figure}
Figure \ref{fig:CaseStudy}a) shows the SoC during each working week for the three drivers.
Stan starts his working week with a fully charged battery and charges at home once the SoC is below $40\%$. Robert also starts his week with a fully charged battery, but charges every time his battery falls below $100\%$. Alfred carefully read the papers of Keil \textit{et al.} \cite{Keil2016,Keil2017} and thus starts at $50\%$ and recharges only if the SoC is below $20\%$. At the end of each working week, all three recharge their EV to their initial SoC and don't move their EVs over the weekend. They all try different charging rates, starting from the standard charging rate C/8 shown in the darkest color, over the brighter C/3, up to the brightest 1C. In the course of one week, all three drive the same range and thus also have the same charge throughput.

In Figure \ref{fig:CaseStudy}b), we show the loss of cyclable lithium after three years for each driver. We clearly see that range anxious Robert's EV shows the highest capacity loss of $10.4\%$, closely followed by standard driver Stan with $9.2\%$. In contrast, aging optimizer Alfred only loses $3.6\%$ of his EV's starting capacity. The different charging rates, colored in brighter colors, hardly affect the battery aging and cause only an increase about $0.2\%$ to $9.4\%$ from C/8 to 1C for standard driver Stan.

We best comprehend the different aging characteristics with the sensitivity analysis of our model shown in Figure \ref{fig:Sensitivity}a). Range anxious Robert and standard driver Stan operate their batteries mostly at high SoCs, contrary to aging optimizer Alfred. Following the red curve in Figure \ref{fig:Sensitivity}a), we see that Robert and Stan are thus mostly on the highest graphite stage, which exhibits nearly the fourfold degradation rate compared to the second graphite stage on which Alfred mostly operates his battery. This is the main cause of accelerated aging that we see in Figure \ref{fig:CaseStudy}b). 

Surprisingly, the charging current plays only a minor role, although according to the yellow line in Figure \ref{fig:Sensitivity}a) it should have the largest impact on the degradation rate. However, because the charging time shortens with increasing current, this dependence is misleading as we also showed with the dashed yellow line in Figure \ref{fig:Sensitivity}a). The presented user-profiles further enrich this analysis and show that degradation depends rather on the charge throughput than the applied charging rate. However, our simplistic model only considers SEI growth as aging mechanism. For lithium plating we would expect a major effect of fast charging rates on aging.

Summarizing the findings of our experimentally validated aging model, we predict that battery aging depends mainly on the operating window, \textit{i.e.} maximum and minimum SoC. Charging current and temperature, in contrast, have only a minor influence.

However, our predictions are based on a homogeneous SEI growth model with some inherent shortcomings.
First of all, DVA and post mortem analysis reveal heterogeneous aging within the cells. Implementing our model in three-dimensional thermal and electrochemical battery models will help to resolve heterogeneous SEI growth \cite{9049261,bolay2021,Chouchane2021}. 
Furthermore, we observe excessive lithium plating in the cell center, which results presumably from thermal hotspots \cite{Storch.2021}. Extending our model for a plating kinetic will not only help to predict these heterogeneities, but also increase our model predictivity for low temperatures.
In the long-term, the locally strong aging will eventually cause "sudden death" of specific cell areas due to electrolyte dry-out. Coupling our SEI-growth model to the percolation model of Kupper \textit{et al.} \cite{Kupper.2018} is a promising approach to capture this effect.
Lastly, DVA reveals an OCV-curve change over time, which will also affect the aging kinetics. Future works can rely on our SEI growth model and refine it for these secondary aging modes to further increase the model predictivity. 

\section{Conclusion}
\label{s:Conclusion}

We applied an electrochemical SEI growth model \cite{VonKolzenberg2020} to predict active lithium loss in 62 automotive grade batteries cycled with 28 different protocols \cite{Stadler.2022}. The simplicity of our model allows us to to parametrize the state of charge, time, current and temperature dependence of the model with only three protocols and the open circuit voltage curve of the anode. The validation with the remaining 25 protocols shows remarkable accordance of predicted and measured active lithium loss with a global root-mean-squared error of 1.28\%.
Thus, our methodology reduces the set of experiments to parametrize a global aging prediction.

For the first time, our so-validated model quantitatively predicts how operating conditions affect battery lifetime. Along three exemplary use cases, we show that the battery operating window, \textit{i.e.} minimum and maximum state of charge, mainly drives aging. In contrast, temperature and current play only a minor role. These insights help in deriving battery design and usage recommendations to prolong lifetime.

Future works can further refine and extend our model for several effects. In particular, post mortem analysis of aged cells with differential voltage analysis reveal lithium plating and heterogeneous aging inside the cells. 
Extending the model for lithium plating is straightforward, because we assume lithium atom mediated SEI growth. To resolve heterogeneous aging, implementing our model in three-dimensional cell simulations is a promising approach \cite{9049261,bolay2021,Chouchane2021}.

\section*{Acknowledgements}
    Lars von Kolzenberg gratefully acknowledges funding and support by the German Research Foundation (DFG) within the research training group SiMET under the project number 281041241/GRK2218. The support of the bwHPC initiative through the use of the JUSTUS HPC facility at Ulm University is acknowledged. This work contributes to the research performed at CELEST (Center for Electrochemical Energy Storage Ulm-Karlsruhe).
    We also thankfully acknowledge Mercedes-Benz AG for funding of the extensive cyclic aging test. We thank Timm Konstantin Groch for his support in the post mortem analysis.

\section*{Conflict of interest} 
    The authors declare no conflict of interest.

\bibliography{literatur_abbr}
\end{document}